# Using Ensemble Monte Carlo Methods to Evaluate Non-Equilibrium Green's Functions, II. Polar-Optical Phonons

**David K. Ferry**

School of Electrical, Computer, and Energy Engineering, Arizona State University, Tempe, AZ 25287-6206; ferry@asu.edu

**Abstract**

In semi-classical transport, it has become common practice over the past few decades to use ensemble Monte Carlo (EMC) methods for the simulation of transport in semiconductor devices. This method utilizes particles while still addressing the full physics within the device, leaving the computational difficulties to the computer. More recently, the study of quantum mechanical effects within the devices, have become important, and have been addressed in semiconductor devices using non-equilibrium Green's functions (NEGF). In using NEGF, one faces considerable computational difficulties. Recently, a particle approach to NEGF has been suggested and preliminary results presented for non-polar optical phonons in Si, which are very localized scattering centers. Here, the problems with long-range polar-optical phonons are discussed and results of the particle-based simulation are used to examine quantum transport in InN at 300K.

---

## 1. Introduction

For almost a century, semi-classical (and classical) simulation of transport in physical systems [1], and particularly in semiconductors [2], has been treated with the Boltzmann transport equation. At low electric fields, a nearly equilibrium situation exists in which the distribution function is a Maxwellian, or a Fermi-Dirac, at a temperature that likely increases little above that of the lattice. This is quite different in semiconductor devices, in which the distribution is well out of equilibrium, which also means that this distribution is basically unknown. Finding this distribution is typically the single most difficult problem. Classically, an alternative approach is to use the computer to completely solve the transport problem with a stochastic methodology, and several methods have arisen to do this, two of which are an integral iteration technique [3,4] and the Monte Carlo method [5]. Today, the most widely used approach in semiconductors is the latter. The ensemble Monte Carlo (EMC) technique uses an ensemble of particles whose propagation is determined in parallel, so that ensemble averages can be computed as a function of time for the various observables of interest [6]. This approach has been widely successful, especially for high electric field transport, and can easily determine the distribution function as well as the various moments of it [7].

More recently, however, quantum effects in semiconductors, and in semiconductor devices, have become important. This arises principally from the small size of the integrated device in modern "chips." As quantum effects have become important, simulation of transport has led to the use of the non-equilibrium Green's functions (NEGF) [8,9], which are analytically extremely difficult. However, recently an approach based upon particle simulation of the NEGF has been proposed (hereafter referred to as I [10]). In this approach, the NEGF are defined in terms of an Airy transform in the direction along the electric field. Results were presented for scattering from the non-polar optical (and intervalley) phonons in Si. Since the non-polar optical mode scattering, as well as the acoustic mode scattering, is highly localized, the assumption of scattering at a point in space can be retained in the quantum simulation.

On the other hand, polar-optical mode scattering is Coulombic in nature [11], arising from the polarization between the two atoms of the basic unit cell in a material such as GaAs, which possesses the zinc-blende crystal structure (other structures may have more atoms, as will be discussed later) [12]. Coulombic scattering is long range, and can give rise to correlations and interference effects in quantum transport. This is particularly true for impurity scattering [13,14] and electron-electron scattering [15]. Such correlations lead to difficulties when these scatterers are introduced perturbatively, as they require an evaluation of the Bethe-Salpeter equation for two- and three-particle Green's function, which are necessary for determination of mobility and transport [16,17]. This raises the difficulty level by a significant amount [18]. However, it has generally been believed that optical-phonon scattering is "phase breaking" [19]. In such a situation, long-range correlations that exist in the phase are destroyed, and there may be no need to proceed to the inclusion of the Bethe-Salpeter equation. This is certainly likely the case for non-polar scattering, which is assumed in I. But, it may also be true for polar-mode scattering, and more work is needed to ascertain whether this is true or not.

In seeking to determine just to what extent phase breaking is important, one can follow two approaches. First, polar-optical phonon scattering can be introduced within the NEGF formalism, which is not difficult with the EMC evaluation of the transport. This would allow direct comparison with semi-classical simulations and highlight whether significant differences (which may arise from correlations) occur. Secondly, one could treat the polar modes as a real space potential, as is done for impurities [17], and examine the occurrence of any correlations. This latter approach suffers from the condition that it is not really known how to handle the polar modes in real space--no one seems to have done it. However, the first approach is doable and will be treated in this paper.

In section 2, the NEGF will be used to evaluate the appropriate self-energies and distribution function with the polar-optical mode interaction. The derivations and discussion of the NEGF were given in I, so only the pertinent equations that properly treat the different matrix element that arises for the polar modes will be presented. The polar-optical mode scattering has certainly been treated semi-classically for low dimensional systems, so that the form of the scattering rates are known for those cases. These will provide a comparison for the quantum situation.

In evaluating the effects of the polar-optical modes on transport, it is desirable to use a material in which a relatively wide range of electric field can be used. In most of the direct gap III-V materials, for example, the conduction band contains satellite valleys at X and L, that are relatively low in energy. This leads to the situation that transfer of carriers from the Γ valley to the satellite valleys begins are low fields of the order of 2-3 kV/cm [7]. On the other hand, the wurtzite phase III-Nitride material InN possesses a small band gap, yet the satellite valleys appear to be very high in energy, so that carriers remain in the Γ valley up to fields of the order of 50 kV/cm. So, InN will be used here to illustrate the role of the polar-optical mode in transport. In section 3, the properties of indium nitride will be discussed rather broadly as the reader may not be familiar with the details of this material. However, the downside of using this material is a dearth of experimental data on high electric field transport, although there are many classical EMC simulations of this transport, as will be discussed in section 3.

The results from the EMC simulation for the transport of electrons at high fields will be presented in section 4. Finally, some discussion and conclusions will be given in section 5.

## 2. NEGF and the Polar Phonon

With the exception of the polar-optical phonon mode, phonons tend to be characterized (and understood) as mechanical vibrations of the atoms in the crystal lattice [20]. The regularity of the lattice imposes quantization on the phonons and insures a degree of coherence in their motion arising from their thermal energy. The polar-optical phonon is different, as it is an electromagnetic response that arises from an ionic contribution to the crystal bonding of the normally covalently-bonded material. In III-V compounds, for example, the atoms in the face-centered cubic lattice are tetrahedrally coordinated, having bonds to the four nearest neighbor atoms. Hence, on average each atom has four bonding electrons. The III-V compounds deviate from this, as one atom has only three electrons and the other has five. Thus, the average atom still has four bonding electrons, but this involves charge transfer from the group V to the group III atom. The amount of average charge on each of the two atoms is known as the *effective charge*, although there are many different forms of effective charge that have been suggested [21,22]. This effective charge leads to a charge dipole existing between the two neighboring atoms that interacts with optical waves in the far infrared, thus creating a difference between the optical dielectric function in the visible ($\epsilon_\infty$) and the static, or low frequency, dielectric function ($\epsilon_s$). This interaction is characterized as a polarization related to the external applied flux density **D** as [23]

$$\boldsymbol{P} = \left(\frac{1}{\epsilon_\infty} - \frac{1}{\epsilon_s}\right)\boldsymbol{D} . \qquad (1)$$

This interacts with the carriers, through the polarization, and results in a long-range interaction due to the Coulombic nature, for the interaction with distant dipoles is also important [24]. While the electric field of the dipole decays as $1/r^3$, the number of dipoles increases as $r^2$. This polarization leads to a scattering interaction of the carriers that may be expressed through the square of the matrix element as (in MKS units) [25]

$$|M|^2 = \frac{m^* e^2 \hbar^2 \omega_0}{8\pi \Omega q^2}\left(\frac{1}{\epsilon_\infty} - \frac{1}{\epsilon_s}\right), \quad (2)$$

which is augmented normally with a function of the Bose-Einstein distribution and an energy conserving delta function [7]. In this latter equation, $m^*$ is the effective mass of the carrier, $\omega_0$ is the radian frequency of the longitudinal polar-optical phonon mode, $\Omega$ is the effective volume of the crystal, and $q$ is the wave number of the momentum exchanged during the interaction. It is this latter momentum exchange that creates a major difference in the non-polar and polar scattering processes.

The retarded self-energy is the product of the retarded Green's function and the retarded phonon Green's function. In most semiconductors, the scattering is weak, and the self-energy can be calculated to lowest order in time-dependent perturbation theory. This retarded self-energy was developed in paper I, and may be expressed as (from equation I.20) [26]

$$\begin{aligned}\Sigma_r(\boldsymbol{k},s,s',\omega) = &\int \frac{d^3q}{8\pi^2 q^2}\sum_{\nu=\pm 1}\frac{eF_0}{2\hbar}\left(N_q + \frac{1+\nu}{2}\right)\\ &\times \int ds_1 \int ds_2\, G_r(\boldsymbol{k}+\nu\boldsymbol{q},s_1,s_2,\kappa)\\ &\times \frac{1}{L^2}\int \frac{dz}{L}\int \frac{dz'}{L} Ai(z-s)\\ &\times Ai(z-s_1)Ai(z'-s')Ai(z'-s_2)\end{aligned} \quad (3)$$

Here, the term [27]

$$eF_0 = \frac{m^* e^2 \omega_0}{4\pi \hbar}\left(\frac{1}{\epsilon_\infty} - \frac{1}{\epsilon_s}\right) \quad (4)$$

arises from the matrix element (2), and the energy conservation as it were lies in the retarded Green's function. This may be compared with an equivalent formulation for the semi-classical polar-optical mode scattering in two-dimensional systems [28], where the delta function and integration over the quantum well wave functions is replaced by the retarded Green's function and integration of the Airy transforms. What is common after integration over the magnitude of q and connection to the initial and final wave vectors is the angular integration [28]

$$I = 2\int_0^\pi \frac{d\vartheta}{q_\pm(\vartheta)}. \quad (5)$$

This denominator can be expressed as

$$q_\pm(\vartheta) = \left[2k^2 \pm \frac{2m^*\omega_0}{\hbar} - 2k\sqrt{k^2 \pm \frac{2m^*\omega_0}{\hbar}}\cos\vartheta\right]^{1/2}. \quad (6)$$

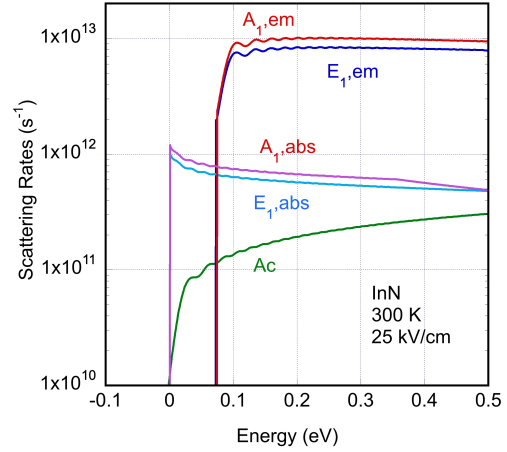

**Figure 1.** The scattering rates for the two polar-optical modes and the acoustic mode in wurtzite InN that enter the retarded self-energy. The two modes are indicated by the labels $E_1$ and $A_1$, which are described fully in section 3 of the text.

This also can be expressed in terms of the energies involved (written simply here for parabolic bands) as

$$q_\pm(\vartheta) = \frac{\sqrt{2m^*}}{\hbar}\left[2E \pm \hbar\omega_0 - 2\sqrt{E(E\pm\hbar\omega_0)}\cos\vartheta\right]^{1/2}. \quad (7)$$

The argument of the overall square root can be written as $A - B\cos\vartheta$, after which the integral (5) can be expressed as [29]

$$I(k) = \frac{2}{\sqrt{A+B}}E\left(\frac{\pi}{2},r\right), \quad (8)$$

where E is a complete elliptic integral and

$$r = \sqrt{\frac{4\sqrt{E(E\pm\hbar\omega_0)}}{2E\pm\hbar\omega_0 + 2\sqrt{E(E\pm\hbar\omega_0)}}} < 1. \quad (9)$$

The pre-factor in (8) becomes

$$\sqrt{A+B} = \sqrt{E} + \sqrt{E\pm\hbar\omega_0}. \quad (10)$$

The scattering rates that go into the retarded self-energy are shown in figure 1. In semi-classical transport, these scattering rates would be determined by the Fermi golden rule, but here they are modified for quantum transport by the Airy functions in (3). These are individual rates that are summed for the imaginary part of the retarded self-energy. Here, it may be noted that there are two dominant polar-optical modes in the wurtzite phase of materials, denoted here by the $E_1$ and $A_1$ nomenclature. These will be described fully in the next section where the properties of InN are described. Here, it is important to continue the discussion of NEGF. It may be noted that, in contrast to the non-polar interactions described in paper I, here there is a tendency for the scattering rates to decrease as the energy increases, which is also seen in the semi-classical case [28,30]. This arises from the effects of (8).

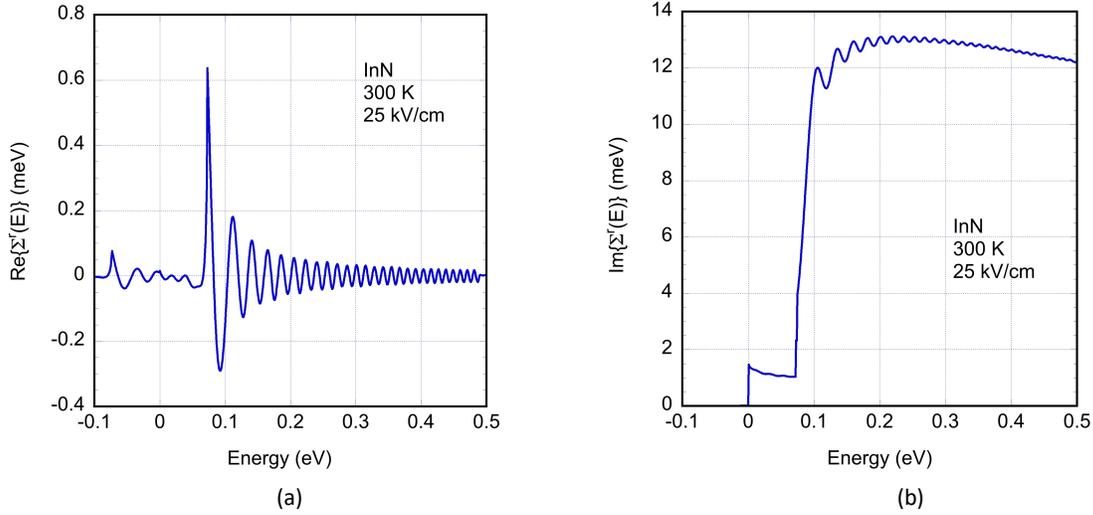

**Figure 2.** (a) The real part of the retarded self-energy at an electric field of 25 kV/cm. (b) The imaginary part of the retarded self-energy at the same electric field.

The real and imaginary parts of the self-energy are given by equation (22) of I, and may be represented in the present case by

$$\Sigma_r(\mathbf{k}, s, \omega) = \frac{eF_0 m^*}{\hbar^2} \sqrt{\frac{\Theta}{2\pi}} I(k) \\ \times \sum_{\nu=\pm 1} \left(N_q + \frac{1+\nu}{2}\right) F(s, \kappa) \quad , \quad (11)$$

where $\Theta = eFL$ was defined in I. The function $F(s, \kappa)$ is given by

$$Im\{F(s,\kappa)\} = Ai'(-y)Bi'(-y) \\ \qquad -yAi(-y)Bi(-y) \quad . \quad (12) \\ Re\{F(s,\kappa)\} = Ai'^2(-y) - yAi^2(-y)$$

The function $\Theta$ enters through the term in $1/L^2$ that precedes the Airy integrals in (3), as

$$L^2 = \frac{L^3}{L} = \frac{\hbar^2}{2m^* eFL} = \frac{\hbar^2}{2m^* \Theta} . \quad (13)$$

In figure 2, the real and imaginary parts of the self-energy, for an electric field of 25 kV/cm, are plotted for the present situation of wurtzite InN. In panel (a), it may be seen that the main peak of the real part actually occurs at the onset of the polar-optical emission process, where the scattering rates greatly increase. Because the onset of the imaginary part is so sudden, the spectral function that is computed from these self-energies is quite sharp with a single major peak near zero energy. This peak is only a few meV in width.

The development of the "less than" Green's functions in the Airy formulation were also developed in I. Nevertheless, it is useful to repeat the final results that were obtained there. One important function is the less-than self-energy, or at least the imaginary part that will appear in the integral equation for the distribution function. This may be written, with the modifications due to the change in matrix elements, as [10]

$$\Sigma^<(\mathbf{k},s,\omega) = \frac{m^* eF_0}{2\pi 3^{1/6} \hbar^2} \sum_{\nu=\pm 1} \left(N_q + \frac{\nu+1}{2}\right) \\ \times \sqrt{\frac{\hbar^2}{2m^*}} \int d^2\mathbf{q}_t \, I(k) \int ds' Ai^2\left(\frac{s-s'}{3^{1/3}L}\right) \quad (14) \\ \times G^<(\mathbf{k} - \nu\mathbf{q}_t, s', \omega - \nu\omega_0) \quad .$$

Here again, the importance of the angular integration appears due to anisotropic nature of polar-mode scattering. As in paper I, this self-energy leads to the integral equation for the distribution function, which will be given in section 4. The imaginary part of the less-than self-energy is shown in figure 3. The significant decrease in this self-energy with the energy itself is a result of the angular integration (8).

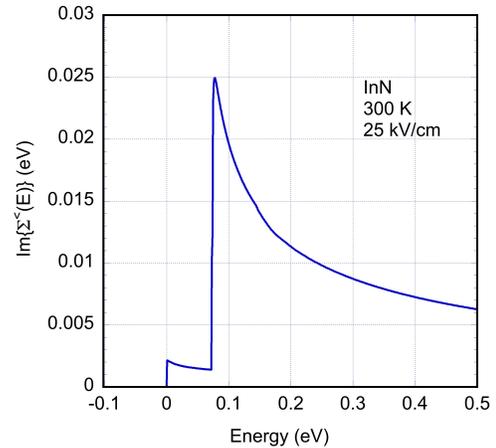

**Figure 3.** The imaginary part of the less-than self-energy for the case of wurtzite InN.

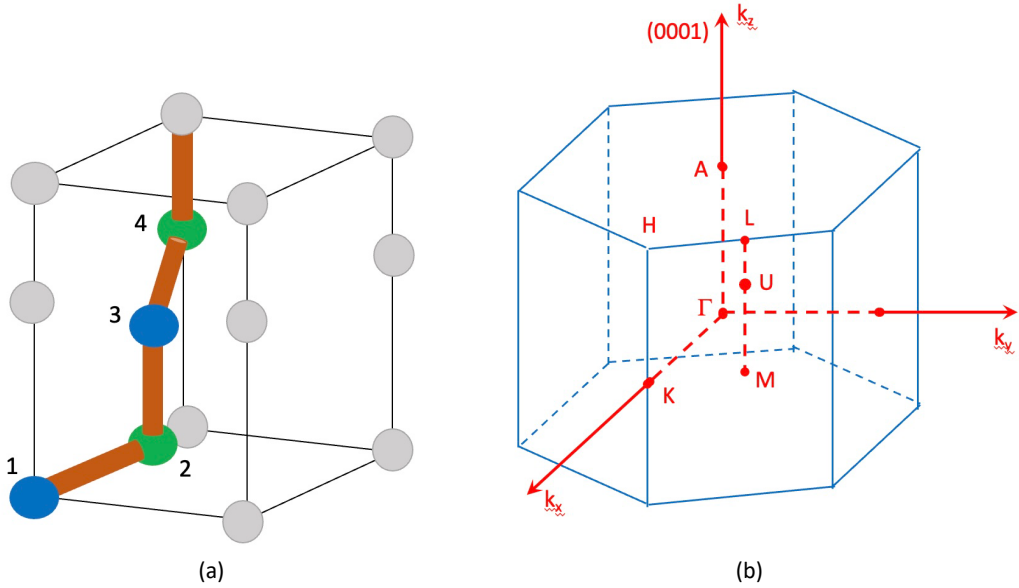

(a) (b)

**Figure 4.** (a) The primitive unit cell for the wurtzite latttice. The In atoms may be considered to be the blue atoms (numberer 1 and 3), while the N atoms are the green atoms (numbered 2 and 4). (b) The Brillouin zone for the wurtzite lattice with key points labeled with the letters. The grey atoms are added merely to outline the shape of other atoms in the wurtzite cell to better illustrate the overall structure..

### 3. Indium Nitride

The InN that is considered here is the dominant wurtzite phase of the crystal. Wurtzite is generally considered to be a relaxation of the zinc-blende structure that occurs when the ionic contribution to the bonding becomes too large [31]. In contrast to zinc-blende however, the Wurtzite primitive unit cell has four atoms, rather than two. This primitive unit cell is shown in figure 4(a). The square top and bottom layers of the cell have edge dimension $a$, and one of these edges forms one leg of the hexagonal bottom plane of the lattice. The height of the cell is $c$ (typically $> a$). The $c$-axis is normal to the hexagonal cell (bottom or top) plane. For example, the two blue atoms (labeled 1 and 3) are the In atoms, while the two green atoms (labeled 2 and 4) are the N atoms. The Brillouin zone for this wurtzite structure is shown in figure 4(b), with various reciprocal lattice points labeled.

The fact that there are four atoms in the basis of the primitive unit cell means that there will now be 12 phonon modes. Three of these are the acoustic modes--one longitudinal and two transverse vibrational modes. There are thus 9 optical phonon modes, not all of which are observable in Raman scattering (the unobservable modes are known as dark modes). The principle modes important for scattering of electrons are the two dominant optical modes with the highest energy and known as $A_1$ and $E_1$ modes. The $A_1$ mode has the two N atoms moving in one direction and the two In atoms moving oppositely with the motion along the (0001) or $z$-axis (the $c$-axis in the figure). The $E_1$ mode is comparable motion but with the motion in the $(x,y)$ plane. There is a lower energy $E_2$ mode that is also in the $(x,y)$ plane, but with one of the atoms (e.g., In) fixed and the other atoms moving relative to them. There are also B modes which are similar to the $E_2$ modes, but with the motion along the $c$-axis, and are often part of the dark modes. The phonon spectrum is shown in figure 5 [32]. In the present study, only the two highest energy $A_1$ and $E_1$ modes are considered. These two phonons have energies of 74 and 75 meV, respectively [33]. These two dominate the carrier scattering and are often driven out of equilibrium, certainly with intense laser Raman studies [34], but also in high electric fields. The non-equilibrium phonons will not be considered here.

An important point is that InN, like other group III-nitrides, possesses an intrinsic permanent polarization of the lattice [35], which differs from the phonons. This static, permanent polarization, as well as the static polarization of the atomic motion, contributes to the band structure, but not to the scattering of the carriers. Scattering arises from oscillatory behavior of the atomic motion, and is handled within normal time-dependent perturbation theory, which is accounted for by the lowest order diagrams in Green's function theory. One advantage of the permanent polarization is that, in heterostructures such as InN grown on GaN, the discontinuity in this polarization across the interface leads to a rather dense two-dimensional carrier density at the interface [36], thus negating the need for actual doping of the material.

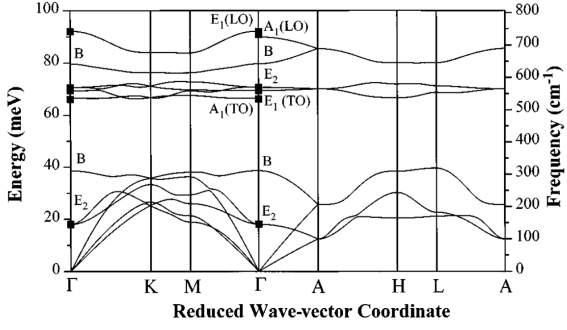

**Figure 5.** Phonon dispersion for the wurtzite lattice. The two highest energy modes, the $E_1$ and $A_1$ LO modes are considered in this work. The modes themselves are discussed in the text. Reprinted with permission from H. Seigl et al. [32], copyright 1997 by the American Physical Society.

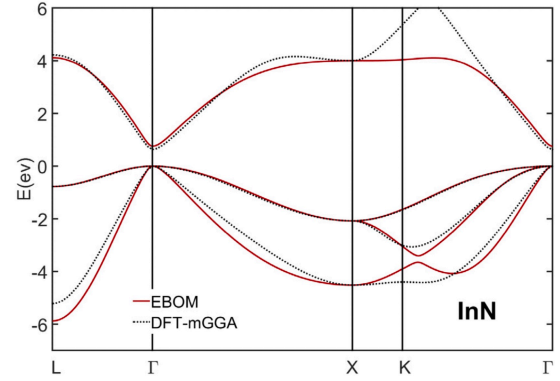

**Figure 6.** The principle energy bands of wurtzite InN around the principle gap at $\Gamma$. Two different methods of calculation are shown, and discussed in the text. Reprinted from Hsiao et al. [37], with permission from Elsevier.

In figure 6, the band structure of InN is plotted for two different methods of computation is shown [37]. One method is a bond-orbital model, which has close similarities to an empirical tight-binding approach but is somewhat simpler in computational effort. The second approach is standard density-functional theory using a GGA functional for exchange and correlation [38]. These calculations give an energy gap of 0.78 eV, and this is used in the present calculations, especially in the non-parabolicity correction terms. Clearly, the principal gap lies at the zone center point $\Gamma$, although a constant energy surface is not strictly spherical. One may note in figure 4(c) that, because $c > a$, the height of the Brillouin zone is smaller than the lateral dimensions. This leads to a slight ellipticity in the constant energy surface and a difference in the values for electron effective mass in the basal plane and normal to this plane. Values used here are $0.055m_0$ in the basal plane and $0.045m_0$ normal to this plane [39,40]. Non-parabolicity of this conduction band has been taken into account by a normal $\mathbf{k}\cdot\mathbf{p}$ model [41].

## 4. The Monte Carlo Approach

The iterative equation for the distribution function was presented in paper I, and may be repeated here from (I.29) as [42]

$$f(s,\omega) = \frac{1}{Im\{\Sigma_r(s,\omega)\}} \sum_{\nu=\pm 1}\left(N_q + \frac{\nu+1}{2}\right) \times \int ds' K(s,s',\omega-\nu\omega_0) f(s',\omega-\nu\omega_0) \quad (15)$$

where

$$K(s,s',\omega) = \frac{\sqrt{3}|M_q|^2 m^*}{\hbar L^2} Ai^2\left(\frac{s-s'}{3^{1/3}L}\right)\left[\frac{\pi}{2} + atanh\left(\frac{\hbar\omega - E_{k,s'} - Re\{\Sigma_r(,s',\omega)\}}{Im\{\Sigma_r(,s',\omega)\}}\right)\right]. \quad (16)$$

It may be noted that the distribution function $f(s,\omega)$ appears both on the left-hand side of (15) and in the final integral on the right-hand side of (15). This makes possible the iteration procedure that is the EMC process. The ensemble of (in this case, $10^5$) electrons is initialized in momentum space as a Maxwellian distribution at 300 K. This initial distribution is used as the starting distribution within the integral. Then, (15) is used to iterate the particles of the distribution until steady-state is reached, although it is quite common to follow the distribution as it evolves in time for transient studies [6]. At various points in the evolution, statistical averages of a quantity $Q$ (average position, momentum, velocity, energy, etc.) is determined by an ensemble average as

$$\langle Q\rangle = \frac{1}{N}\sum_i Q_i, \quad (17)$$

where the sum runs over the values for each of the $N$ particles.

The difference between the semi-classical EMC and that used here lies in the acceleration time step. In the semi-classical EMC, each electron has an acceleration time step whose statistical probability is determined by the temporal scattering rate for a carrier, and velocity and distance are determined from this time step. Here, however, the acceleration lasts for a distance traveled rather than a time, and this distance is determined by the distance Δ$s$ that is determined from the $Ai^2$ term in the integral of (15). Naturally, the time Δ$t$ and the distance Δ$s$ are related to one another. This relationship in nonparabolic bands takes the form [10]

$$\Delta t = \sqrt{\frac{2m_c}{eF}|\Delta s|\left(1 + \frac{E}{E_G} - \frac{eF\Delta s}{2E_G}\right)} \qquad (18)$$

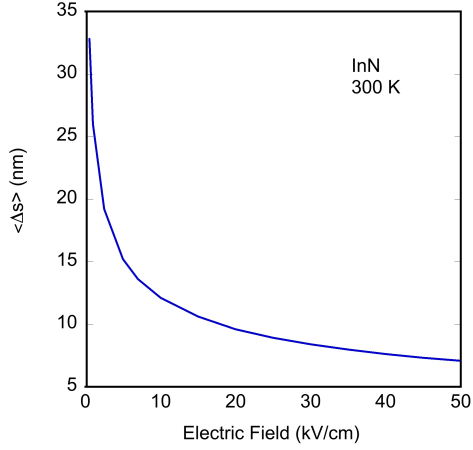

**Figure 7.** The average distance traveled during acceleration of the particles by the electric field.

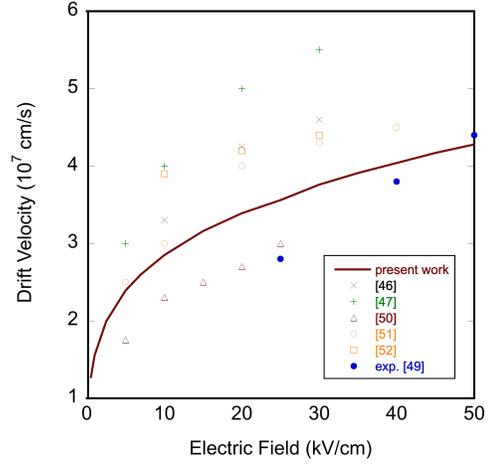

**Figure 8.** The computed velocity-field values for the NEGF simulation (solid curve). Various values for semi-classical simulations and some experimental data are shown for comparison (references are indicated for these).

From this, changes of the momentum, velocity, and the time step are determined, which is basically the reverse procedure from that used in the semi-classical EMC. This procedure is fully detailed in paper I. In figure 7, the average Δ$s$ is plotted as a function of the applied electric field. As this depends upon the field through the definition of $L$ in (13), and various materials parameters, this plot differs from that of the earlier paper, as this one is for InN and polar-optical phonon scattering.

The computed drift velocity of the ensemble used here is shown in figure 8. This is determined by averaging over the $10^5$ particles and the 200 iterations of the EMC procedure, as described in I [10]. Thus, this is a time and ensemble average. Several semi-classical EMC simulations have appeared in the literature, and these have given a wide range of values for the electric field value for the peak of the velocity-field curve, which traditionally signals the onset of carrier transfer to higher-lying conduction band valleys. Values that have appeared in the literature include 25 kV/cm [43], 30 kV/cm [44,45,46], 35 kV/cm [47], and 40 kV/cm [48]. Our own experimental studies, using single-particle Raman scattering in the presence of the high electric field [49], suggests that the peak velocity occurs at fields of the order of 75 kV/cm. But, it must be stated that this all depends upon many material parameters whose values are not fully determined. For that reason, as mentioned earlier, fields above 50 kV/cm have not been used in this work. However, several estimates of the velocity from these semi-classical simulations are also plotted in figure 8 for comparison to the present velocity values [46,47,50,51,52]. The experimental data from [49] is also shown. The first thing one observes is that the semi-classical EMC data are quite varied with no real agreement between them. Our own semi-classical EMC, used in [49] gives some agreement with the experimental data shown in the figure. What that calculation included, which is not present in the current work, was scattering from impurities and defects/dislocations [53,54], both of which greatly reduce the low-field mobility and therefore affect the high-field values of velocity. At present, defects and dislocations in particular affect all group III-nitrides.

The distribution functions for three values of the electric field are shown in figure 9. These have quite similar shapes, although there is significantly more streaming to high energies at the higher values of the electric field. What is also significant is the two plateaus that appear in each curve. The breaks in the curves seems to be at the polar-optical phonon energy and twice this energy. This appears to be unusual, and will be discussed in the discussion of the next section. The peak amplitude of each the three distributions is similar, although somewhat lower at high fields due to the increased streaming. While the values for each distribution are relative to one another, each is normalized to contain the same number of particles.

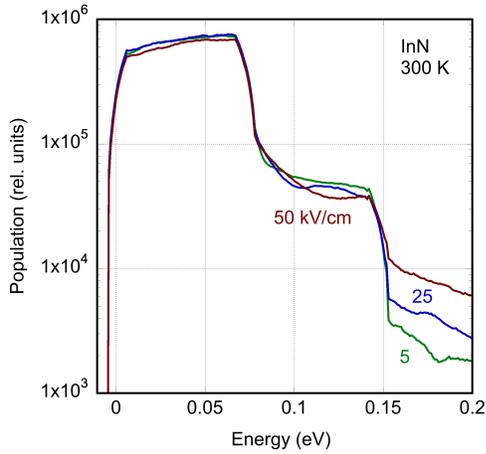

**Figure 9.** The carrier distribution function for three different values of the applied electric field.

## 5. Discussion

Probably the most obvious point is that the semi-classical simulations appear to be all over the place (figure 8), and there is no coherence within the group of simulations used in the figure. While this might be due to the range of values that exist for the material properties, nearly all of these simulations use similar values for the phonon energies and the effective mass of the carriers, so the wide range of results is surprising, but perhaps says more about the lack of experimental data with which to compare the simulations.

Nevertheless, it is apparent that polar-optical phonon modes can be treated within the Airy function approach. But, this avoids answering the question as to whether the assumption of phase breaking in optical phonon scattering is a warranted assumption. The distribution function (figure 9) has two major steps/breaks which appear to occur at the onset of phonon emission by the two high-energy modes, and also at twice this energy. In paper I, something similar occurs, as there is a noticeable break just above the high-energy phonon emission energy and another weaker break at twice the energy (figure 9(b) of I). The breaks are much weaker in the Si case, as the optical scattering process in Si increases with energy, while the polar-optical scattering process a large decrease with energy can be seen in figure 3. This would enhance the carrier streaming, in response to the electric field, up to the threshold for phonon emission and likely be apparent at twice the phonon energy as emission of a first phonon leaves the electron at the maximum probability of emission in figure 3 and a high likelihood of a subsequent phonon emission, thus leading the appearance of the two strong breaks.

In neither of the two cases that have been examined (Si and InN) have two-phonon processes been included in the physics. Such a process usually occurs with a higher-order diagram, and these have not been treated; the main point in the phase breaking argument is that these can be ignored. But, one must certainly consider the possibility that the two strong breaks in the distribution might well signal that there is some remaining correlation/coherence in the particle wave, which would indicate the lack of complete phase breaking. It is not apparent that going to a consideration of the integral equation arising from the Bethe-Salpeter equation would answer the question, especially as this latter equation has not been treated by EMC techniques previously. But, typically such interferences are observed in real space, even with particle methods [13,14]. It may be concluded that the assumption of phase breaking in optical phonon interactions needs further study, especially in the case of polar-optical phonon scattering.

While the previous paper [10], and the present work have treated three-dimensional semiconductors, it should directly translate to reduced dimensionality that arises in modern materials and devices. In this work, it may be noticed that the transverse directions appear to become two-dimensional, as may be noticed in (16). Hence, in a quasi-two-dimensional system, the transverse dimension would tend to one dimension, and in a quasi-dimensional system, the transverse dimension would be quasi-zero dimensional. But such systems have been studied for many years [19], and this would entail no complications to the method developed in this work. Indeed, transverse quantum states for such reduced dimensional systems would appear in the retarded (and advanced) Green's functions in (3), as Green's functions are known to be summations over quantum states [9].

All the work here, and in I, has been at rooom temperature. As devices operate at different temperatures, it is reasonable to estimate how this will affect the present results. First, at lower temperatures, the phonons will be less excited according to the Bose-Einstein statistics, so the polar-optical phonon will be less important. However, even at a lattice temperature of 300 K, the actual carrier temperature in the presence of these high electric fields is much higher [7]. Operating devices suffer from raised lattice temperatures due to the need to dissipate the energy input from the electric field and currents, and this plus the elevated carrier temperature can lead to non-equilibrium phonon populations as well. This then becomes a much more complicated many-body problem in which both carrier and phonon Green's functions are involved. While such an investigation would be interesting, it goes beyond the introduction that is considered here.